\documentclass[preprint]{aastex}

\newcommand{\masr}{mas yr$^{-1}$}
\newcommand{\cdeg}{^\circ}
\newcommand{\jb}{Jy beam$^{-1}$}
\newcommand{\mjb}{mJy beam$^{-1}$}
\newcommand{\kms}{km s$^{-1}$}
\newcommand{\sqt}{$\sqrt{2}$}
\newcommand{\emu}{pc cm$^{-6}$}

\slugcomment{To appear in the April 2003 Astronomical Journal}

\shortauthors{Walker \& Anantharamaiah}
\shorttitle{Recombination Line in 3C~84}

\begin{document}


\title{\vspace{1cm} A VLBA SEARCH FOR A STIMULATED RECOMBINATION 
LINE FROM THE ACCRETION REGION IN NGC~1275}

\author{R. C. Walker}
\affil{National Radio Astronomy Observatory\footnote{The National
Radio Astronomy Observatory is a facility of the National Science
Foundation, operated under cooperative agreement by Associated
Universities, Inc.}, Socorro, NM 87801}
\email{cwalker@nrao.edu}

\author{K. R. Anantharamaiah \footnote{Deceased}}
\affil{National Radio Astronomy Observatory, Socorro, NM 87801 and 
Raman Research Instutute, Bangalore, India}


\vskip 1cm

\begin{abstract}

The radio source 3C~84, in NGC~1275, has a two sided structure on
parsec scales.  The northern feature, presumed to be associated with a
jet moving away from the Earth, shows strong evidence for free-free
absorption.  The ionized gas responsible for that absorption would be
a source of detectable stimulated recombination line emission for a
wide range of physical conditions.  The VLBA has been used to search
for the H65$\alpha$ hydrogen recombination line.  The line is only
expected to be seen against the northern feature which contains a
small fraction of the total radio flux density.  This spatial
discrimination significantly aids the search for a weak line.  No line
was seen, with upper limits of roughly 15\% of the continuum over a
velocity range of 1486 \kms\ with resolutions up to 6.6 \kms.  In the
absence of a strong radiation field, this would imply that the
free-free absorbing gas has a wide velocity width, is moving rapidly
relative to the systemic velocity, or is concentrated in a thin, high
density structure.  All of these possibilities are reasonably likely
close to an AGN.  However, in the intense radiation environment of the
AGN, even considering only the radiation we actually observe passing
through the free-free absorbing gas, the non-detection is probably
assured by a combination of saturation and radiation damping.

\end{abstract}


\section{Introduction}

The radio source 3C~84 is associated with NGC~1275, the dominent
member of the Perseus Cluster.  The radio source has a complex
structure on parsec scales that has been an object of study with VLBI
since the very early days of that technique (see Walker et al. 2000
and references therein).  There is a bright core region, compact at
high frequencies, which is thought to be close to the central engine
of the AGN.  Extending to the south, currently to about 15
milliarcseconds (mas), is a complex structure clearly related to the
jet seen on larger scales, but showing a morphology reminiscent of a
radio lobe.  That structure has been moving outward at approximately
0.3 \masr ~and its size extrapolates to zero at about 1959 when 3C~84
started to increase in flux density from below 10 Jy to values in
excess of 50 Jy reached in the 1970's and 1980's.  The flux density
has been decreasing lately, but 3C~84 is still one of the brightest
compact sources in the sky.  HI absorption is seen against the nuclear
region at both the optical systemic velocity of $v=5260$ \kms\
\citep{dV91} and at the velocity of a possibly infalling system at
about $v=8200$ \kms.  VLBI observations of this absorption have
provided information on the intervening clouds \citep{M02} which are
thought to be located along the line-of-sight far from the nucleus.

In 1993, a structure extending about 8 mas north of the core was found
in early Very Long Baseline Array \citep[VLBA;][]{N94}
observations at 8.4 GHz \citep{W94} and in Global VLBI Network
observations from 1991 at 22 GHz \citep{V94}.  By assuming that the
northern and southern features originated in the core at the same
time, a simple relativistic beaming model gives an intrinsic jet
velocity of about $0.4c$ and an angle to the line of sight of about
$40\cdeg$ \citep[][updated for $H_o=75$ \kms Mpc$^{-1}$, the value
that is used in this paper]{W94}.  The assumption of a common start
time is reasonable given the flux history of the source and the
measured velocities.  The northern feature was much brighter at the
higher frequency, suggesting that the radiation is free-free absorbed.
The absorption is only seen against the counterjet.  The near-side jet
is not absorbed at all and the core is at most weakly absorbed;
synchrotron self-absorption may be responsible for the inverted
spectrum.  Therefore, the absorbing material is most likely located on
parsec scales ($1\ {\rm pc} = 3\ {\rm mas}$), probably associated with
the presumed accretion disk that feeds a central black hole.  The
implications of the observed absorption for disk structure are
discussed by \citet{L95}.

The counterjet absorption was studied in detail with near-simultaneous
VLBA observations at many frequencies in 1995 by \citet{W00}.  It was
also seen on somewhat larger scales at lower frequencies by
\citet{S98}.  These observations left little room to doubt that
free-free absorption is the explanation of the observed spectrum of
the counter jet.  They also were able to determine the two-dimensional
distribution of the absorption over the region of the counterjet.
That structure was found to be dominated by a radial gradient
decreasing with core distance.  If the gradient is fitted with a power
law, the index is a bit above $-2$.  But a power law is not an
especially good description of the observed gradient.  If a
temperature of $10^4$ K is assumed, the absorption implies a gas with
an emission measure (EM) of about $5\times 10^8$ \emu\ at a projected
distance of 2.5 pc from the core.

The theory of radio recombination lines predicts that weak maser-like
emission can occur over almost the entire radio frequency range due to
non-LTE effects \citep[see, for example,][]{S76}.  The strength of the
lines, determined by the degree of inversion of the population levels,
is a sensitive function of density and temperature.  Calculations of
the expected strength of recombination lines for gas parameters
implied by the free-free absorption and geometry in 3C~84, in a benign
radiation environment, showed that there was a significant chance of
being able to detect such lines.  An observation of such a line would
provide significant constraints on the physical conditions in the
absorbing gas.  The calculations indicated that the strongest lines
would be those in the general vicinity of 20 GHz.  

The observation of recombination lines from the counterjet region of
3C~84 is tricky.  The counterjet only provides a small fraction of the
total flux density, so, even if the optical depth against that region
is moderately high, the effect on the total power spectrum would be
small.  This would place severe requirements on the bandpass
calibration of observations that don't resolve the source.  We have
used the VLBA, along with one antenna of the VLA, to attempt to detect
the H65$\alpha$ line ($\nu_{rest}=23.404284$ GHz) in 3C~84.  With the
resolution of the VLBA, the region where the line is expected to be
seen can be separated spatially from the rest of the source, relaxing
the bandpass calibration requirements.  And the fact that the
recombination line is not expected to be seen against the bright parts
of the source allows those regions to be used to do an accurate
bandpass calibration using a method developed for these observations.

The calculations of expected line strengths that encouraged us to make
the observations reported here ignored the possible effects of the
intense radio radiation from the jets.  It turns out that those
effects can be severe.  \citet{S78} showed that a strong background
radio source can alter the level populations and saturate the
recombination line maser.  With saturation, the amount of
amplification is limited by the pump rate, so the line strength can be
much lower than would be calculated without taking saturation into
account.  The conditions in 3C~84 are such that the system is likely
to be at or near saturation.  Even more serious, \citet{S78} pointed
out that radio recombination lines will be severely broadened by
radiation damping in the presence of a strong radio source.  Inserting
numbers appropriate for 3C~84 into equation 24 of \citet{S78} gives a
line width of thousands of \kms.  This would make the line
unobservable.  Therefore detection of a line would imply either that
the absorbing region is at least twice as far from the radio
source than implied by the geometry described above, or that something
is wrong with the physics.  Likewise, a non-detection, unfortunately,
provides no interesting constraints on the physical conditions in the
3C~84 accretion region.

\section{The Observations}

On 1998 September 18, 3C~84 was observed in left circular polarization
on the VLBA and one antenna of the VLA.  The frequency range $22937.08
- 23051.08$ MHz was spanned using 8 slightly-overlapped, 16 MHz
baseband channels.  The correlator provided 32 spectral channels for
each baseband channel.  After throwing out edge channels from each
baseband, the final data set retained a total of 224 spectral
channels.  The frequency was chosen so that the H65$\alpha$
recombination line, at the systemic velocity, would be red-shifted to
the center of the fifth baseband channel.  The total velocity range in
the final spectra is $4618$ to $6104$ \kms, or $-642$ to $+844$ \kms
relative to the systemic velocity.  The individual channel
velocity width is 6.63 \kms.  The VLA electronics system limited
coverage at that station to baseband channels $3-7$.


The data were processed in AIPS.  Initial amplitude calibration was
done in the usual manner based on continuously measured system
temperatures and on gain values provided by the VLBA staff.  Two
calibrator scans at the 22.2 GHz frequency where VLBA gains are
measured confirmed that the calibration at 23.0 GHz did not need a
significant frequency dependent adjustment.  But the flux scale was
adjusted upward significantly when the gain normalization was confined
to the 3 best calibrated antennas.  Even that left the total flux
density of 3C~84 about 12\% below the $15.5 \pm 0.5$ Jy measured by
contemporaneous VLBA pointing observations.  As a compromise between
believing the interferometer calibration and trying to match the total
flux density of 3C84, an upward adjustment was made to cut the
difference in half.  The total flux scale adjustment relative to the
blind a priori calibration was 28\%.

\begin{figure} 
\epsscale{0.60} 
\plotone{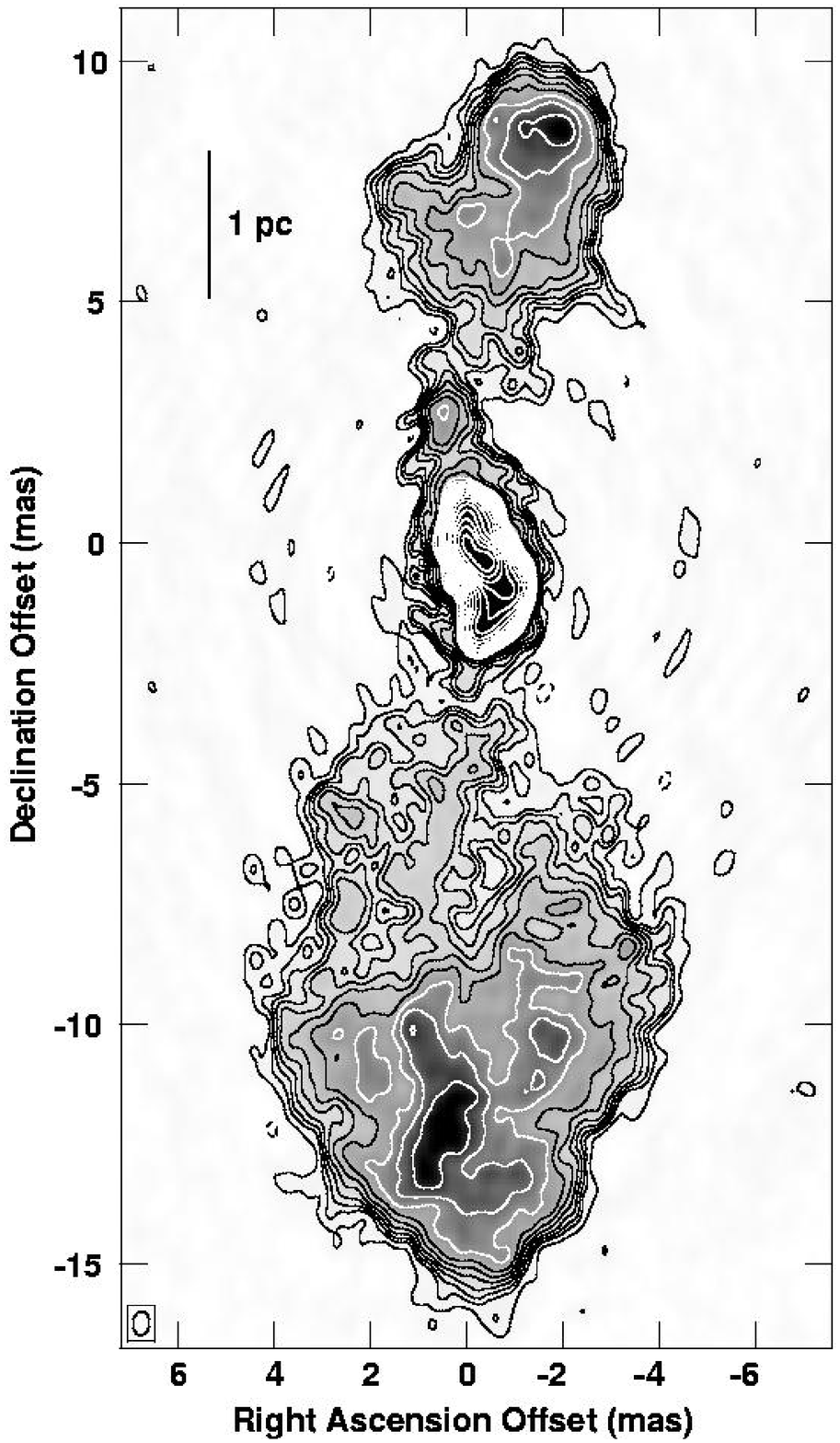} 
\caption{Continuum image of 3C~84 at 23 GHz based on data from
observations on 1998 September 18.  The contour levels start with -1,
1, 2, 2.83, 4, 5.66 m\jb\ and increase from there by factors of \sqt.
The CLEAN convolving beam is $0.49 \times 0.35$ mas elongated in
position angle $-4.3\cdeg$ as indicated by the symbol in the lower
left corner.  The gray scale shows the same image as the contours.
The absolute position is not determined by these self-calibrated
observations, so only offsets from the brightest feature are shown.
\label{image}} 
\end{figure}

\placefigure{image}

For the spectral calibration, a high quality image was required based
on all of the data.  For this, the data were fringe fitted and then
bandpass calibrated using a single bandpass for the full time of the
data set.  Many iterations of self-calibration and imaging
produced the final image shown in Figure~\ref{image}.  The
off-source RMS noise level is 0.29 m\jb.  The peak flux density in the
image is 1.9 \jb\ and the integrated flux density is 14.6 Jy.  For the
northern feature alone, the peak is 42 \mjb and the total is 1.20
Jy. The resolution (convolving beam) is $0.49 \times 0.35$ mas
elongated in position angle $-4.3\cdeg$.  Comparison with the image
from 1995 October \citep{W00} shows good correspondence between even
fairly minor features.  But there have been noticable changes
including expansion away from the core in both the north and south.

By a somewhat circuitous scheme using time dependent bandpass 
calibrations, we effectively self-calibrated each spectral channel
independently using the CLEAN component model from the image described
above.  The model was then subtracted from the data and images were
made of each channel.  We call these the ``difference images''.
These images turn out to be essentially pure noise and were
not deconvolved.

\begin{figure} 
\epsscale{0.63} 
\plotone{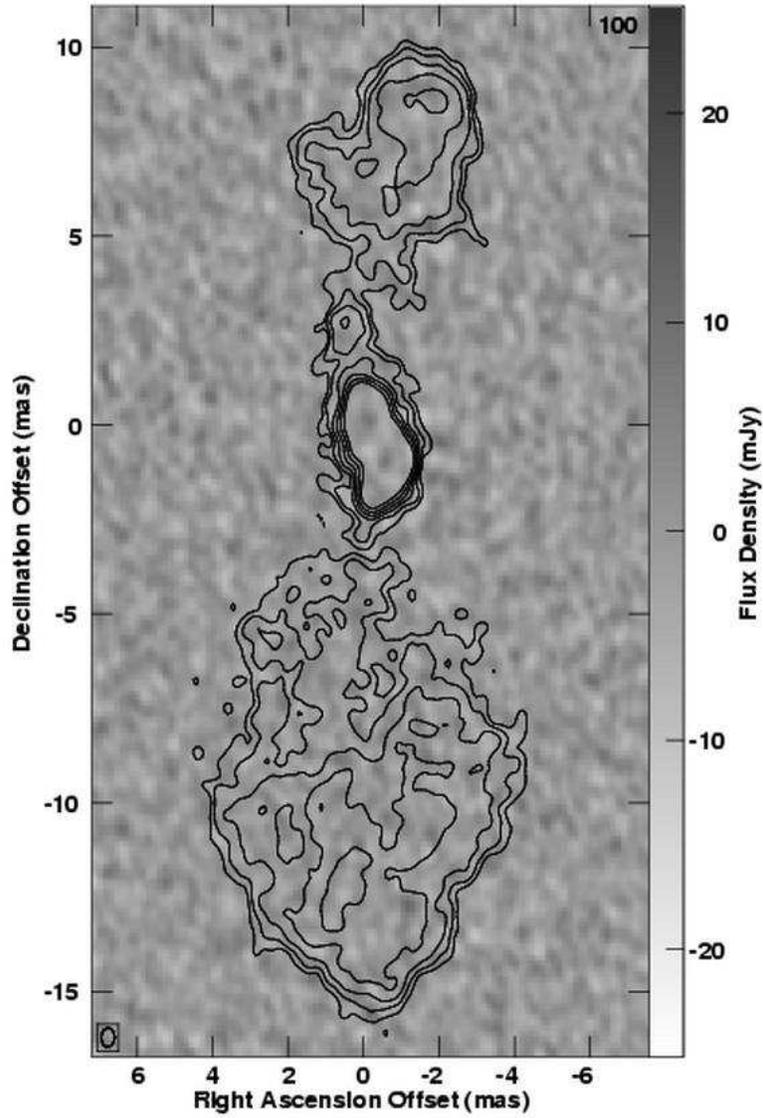} 
\caption{Difference image for Channel 100 of 224 with the -8, -4, -2,
2, 4, 8, 16, 32, \& 64 m\jb\ contours from Figure~\ref{image}
superimposed.  The grey scale range is indicated by the scale on the
right.  The image is made from UV data from which the continuum image
has been subtracted.  No signal is detected in this or any other
channel.  The resolution of the difference image is $0.49 \times 0.35$
mas elongated in position angle $-3.3\cdeg$ as indicated by the symbol
in the lower left corner.
\label{imnoise}} 
\end{figure}

\placefigure{imnoise}

As a representative example of the difference images,
Figure~\ref{imnoise} shows a grey scale display of channel 100.  A
coarse contour plot of the continuum image is overlayed for
orientation.  This channel is typical in that the difference image
looks like noise with no distinction between on and off source
regions.  This means that the data and model match very well with any
residuals lying well below the noise level.  This is not surprising
given the fact that the continuum image, which is based on all 224
channels, has a noise level about 10 times below that of the
individual channel images.

In the procedure described above, the data were being forced to match
as closely as possible the predictions of the image, which in turn had
been made from the data.  But always an assumption was made that the
image is the same in all channels.  Any violation of that assumption
should show up in the final images, which are essentially the
residuals from the model on a channel by channel basis.  If the source
structure were different in a channel, as it would be if a spatially
variant spectral line were present, the difference image should show
the spatial structure of the differences.  Actually, this is only true
to first order because some of the differences will be absorbed into
the channel dependent calibration factors.  To test the magnitude of
this effect, we modified the data for one channel to mimic the
presence of a 10 mJy additional point source somewhere in the middle of
the northern feature.  We then processed the data in the same way that
we processed the real data.  The point source appeared in the
difference image, reduced in amplitude by 23\%.  That indicates
that the self-calibration can scatter about a quarter of the real
difference into calibration parameters.  Note that a spatially invarient
spectral line would just appear as a gain offset in the affected channels
and would not appear in the difference images.

In addition to processing the data as described above, we made
difference images based on a more tradional bandpass calibration using
0528+134.  The only channel dependent calibration derived from the
3C~84 data itself was the removal of linear phase slopes with a fringe
fit.  The difference images derived in this way for all channels were
indistinguishable from noise, but with noise levels about twice as
high as in the channel images based on channel by channel
self-calibration.  This exercise adds confidence that we have not
self-calibrated away a line and that there is no spatially invarient
line covering the whole source.  But lower limits on line strength, even
given the reductions we estimate due to the self-calibration, are
provided by the images based on self-calibration of each channel.
Basically the bandpass calibration provided by self-calibration on
the source itself is significantly better than that provided by the
data on the calibrator.

\section{Results}

No recombination line was detected in any of the images in either
emission or absorption.  The limits presented here are from the
difference image cube based on data with each channel separately
self-calibrated.  The rms noise in the difference image is 3.0 \mjb~
with 0.5 MHz spectral resolution essentially independent of position.
For lower spectral or spatial resolution, up to factors of a few, the
noise decreases about as expected.  

\begin{figure}
\epsscale{0.50}
\plotone{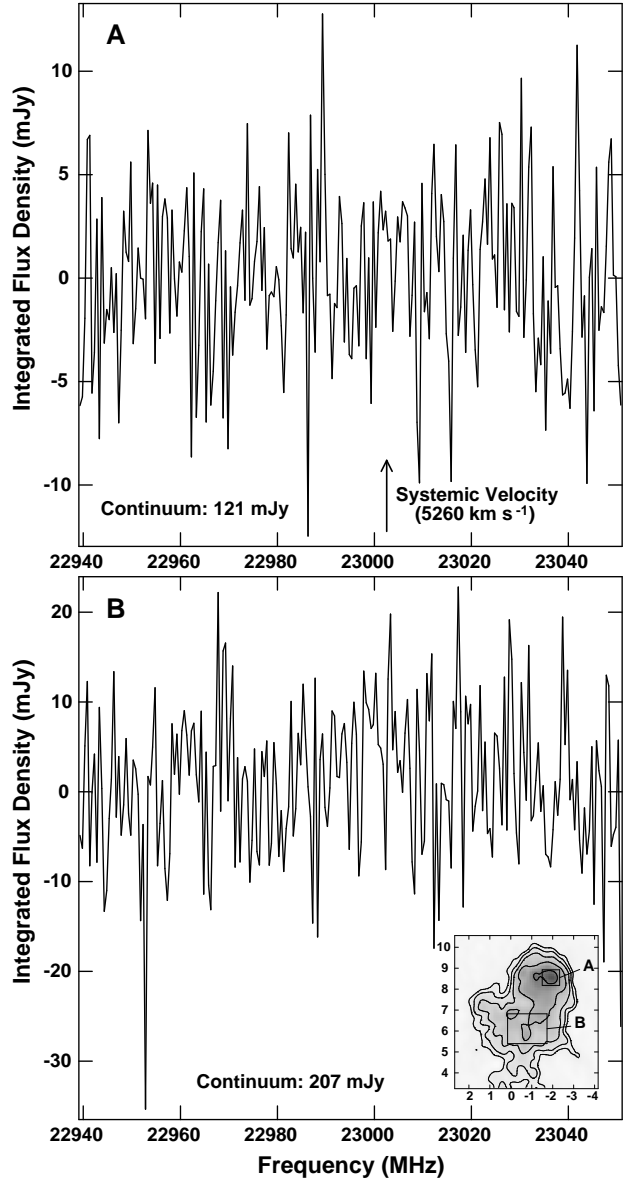}
\caption{Difference spectra for two regions of the
northern feature of 3C~84.  Each spectrum is the integrated flux
density in the region of the difference image.  The inset shows the
northern feature from the continuum image with the the regions for the
two spectra marked and labeled.  These are just meant to be
representative spectra.  The total continuum flux density within each
region is written on the panel.  The largest spikes are at the edges
of baseband channels and therefore are suspect.  The spectra
cover the velocity range 6104 to 4618 \kms.
\label{spnoise}}
\end{figure}

\placefigure{spnoise}

Two example spectra from the difference image cube are shown in
Figure~\ref{spnoise}.  Each is the spatially integrated difference
flux density spectrum for one of two regions of the northern feature
whose locations are shown in the inset.  Region~A includes the peak of
the northern feature while Region~B is closer to the core where a
stronger line might be expected based on the radial gradient of the
free-free absorption found by \citet{W00}.  For Regions~A and B, the
integrated flux densities in the continuum image are 121 and 207 mJy
respectively while the spectra have one sigma noise levels of 4.1 and
8.5 mJy.  The three sigma upper limits to the line-to-continuum ratio
are 0.10 and 0.12.  Adjusting for the possible degradation of flux due
to self-calibration, discussed earlier, these results suggest that any
line is less than about 15\% of the continuum.  This is the value that
will be used in the discussion below.  Note that this limit required
some spatial integration.  On a point by point basis, the three sigma
limit to the line-to-continuum ratio is not much better than 0.5.

Note that the difference image noise level is about $2 \times 10^{-4}$
of the total continuum flux density.  Any attempt to make an
equivalently sensitive search for the recombination line with a single
dish or an interferometer that could not resolve the northern feature
would have required bandpass calibration to better than a part in
$10^4$, a difficult limit to achieve.

\section{Discussion}

As noted in the Introduction, the radio continuum observations give
good evidence that the parsec scale structure of 3C~84 includes an
opposing pair of jets inclined at about $40\cdeg$ to the line-of-sight
with significant amounts of ionized gas in the region between the
jets.  The ionized gas is most likely associated with the accretion
disk that would be there in most AGN models.  The projected distance
from the northern feature to the core is about 2.5 pc so, with the
above geometry, the region of the jet that emits the absorbed
radiation is about 5 pc from where the radiation passes through the
disk on the way to the Earth.  That is a small enough distance
that, as also noted in the Introduction, gas at the disk would be
subject to a sufficiently high radiation intensity that radiation
damping would be expected to broaden the H65$\alpha$ recombination
line to thousands of \kms, preventing detection in our observations.
Our non-detection is consistent with this expectation.

If a line had been detected, the lack of excessive radiation damping
would constrain the gas to lie beyond some minimum distance from the
radio emitting regions.  The actual distance limit depends on which
portions of the overall radio source illuminate the absorbing gas.
The bare minimum illumination is from the northern feature, the
feature for which the free-free absorption is seen.  If that is the
only radiation illuminating the absorbing gas, the separation between
the northern jet and the absorbing region would have to be at least 11
pc for the radiation damping line width to be below the approximatly
500 \kms\ upper limit for what could have been detected in our
observations.  That is over twice the distance between the northern
jet and the disk in the geometry derived from the continuum
observations.  It is a bare minimum distance, because both the
southern feature (near side jet) and the core region most likely also
illuminate the absorption region.  With this additional illumination,
the minimum distance between the emitting and absorbing regions would
need to be much higher, up to about 50 pc if all regions illuminate
the absorbing gas.  Thus if a line had been observed, and the physics
is right, the absorbing region would have to be sufficiently far from
the various radio sources that the idea that it is associated with the
accretion disk would be wrong.  That, in turn, would leave us
struggling to explain why only the northern feature is free-free
absorbed.

\placefigure{recomb10000}

The free-free absorption of the northern feature gives the EM, for any
assumed temperature, of the ionized gas along the line-of-sight
\citep{W00}.  The EM is a function of the electron density and the
absorbing region thickness so its value constrains those parameters.
The upper panel of Figure~\ref{recomb10000} shows the constraints for
an assumed temperature of $10^4$ K.  Constraints for a range of values
of the EM are shown representing the range observed over different
positions along the northern feature. Two other limits are also shown.
The horizontal line is at 2.5 pc, the projected core distance of the
centroid of the counterjet.  This is a rough upper limit for the
thickness of the absorbing region if it is associated with the
accretion disk.  For the observed emission measures, this thickness
limit constrains the density to values above about $10^4$ cm$^{-3}$.
A second limit, given by the lack of significant Thomson optical
depth, is also shown.  Given the observed free-free absorption, the
Thompson optical depth would only be important at very large sizes and
low densities.  This could become an issue if the absorbing region is
actually far enough from the AGN core to avoid severe line broadening
by radiation damping.

If saturation and radiation damping were not a problem, the
non-detection of H65$\alpha$ could be used, along with the measured
EM, to place some constraints on the physical conditions of the
free-free absorbing gas.  For completeness, we will show those
constraints.  The constraints would be of interest if the absorbing
region is actually farther from the continuum radiation source than we
believe or if for some reason the radiation damping is not as severe
as we believe.

Using the measured EM and the assumed temperature, the optical depth
of maser-like recombination line emission that would be expected from
the absorbing gas can be calculated for a range of values of density
and line width.  The lower panel of Figure~\ref{recomb10000} shows the
results of such calculations for an emission measure that is midway
between the values obtained from the continuum absorption.  The
calculations are based on departure coefficients very similar to those
of \citet{SB79} and are for the case when external radiation does not
affect the level populations or line width.  Optical depths are
plotted for several different line widths.  The figure also shows the
upper limit to the line-to-continuum ratio of 0.15 that was derived
above from the data.  In the absence of saturation or radiation
damping, Figure~\ref{recomb10000} shows that the recombination line
non-detection would imply that the absorbing gas must have an
appropriate combination of high density or high line width, presumably
due to turbulence or other dynamic effects.  Or it could be at a
velocity outside the observed window.  Note that the thermal line
width for the assumed temperature is lower than the smallest velocity
width shown.

To put the velocity range and line widths observed into some
perspective, consider that the orbital velocity at about 3.3 pc from a
$4\times10^8 M_\sun$ black hole is about 700 \kms, or about half of
the total observed velocity range.  That mass is clearly uncertain but
is what was estimated by \citet{WH01} using the correlation between
black hole mass and bulge velocity dispersion.  The 3.3 pc distance is
just the observed offset of 2.5 pc deprojected along the disk for the
geometry described earlier.  That orbital velocity shows that it is
possible for gravitationally bound material in the region of interest
to be moving fast enough to produce line widths high enough to
explain our non-detection.  But it also shows that, if the absorbing
material is moving with the disk, it could be observable.  This would
depend on the reasonable assumptions that the disk is moving
transversely where it crosses the sight lines to the far-side jet, and
that it has an internal velocity spread that is much smaller than the
orbital velocity.

The Broad Line Region (BLR) line widths in NGC~1275 are of the order
of our total frequency span \citep[see, for example,][]{N99}.  A
recombination line that wide would both not be sufficiently intense to
observe, according to Figure~\ref{recomb10000}, and would be washed
out by our primary data reduction technique.  But typical BLR sizes,
determined from reverberation mapping, are much smaller than the
region probed by the sight lines through the free-free absorbing
material.  Therefore the BLR line widths would only be a problem if
the high velocities are maintained well beyond the region that emits
the optical lines.


The most rudimentary models of accretion disks predict temperatures at
distances such as those of interest here (few parsecs) that are too
low to imply the presence of ionized gas.  But the free-free
absorption shows that such gas is present.  Other lines of evidence,
based on attempts to understand optical lines, also suggest that
ionized gas is present \citep[see][and references therin]{C99}.  It is
possible that such gas is not in the plane of the disk, but rather in
an atmosphere or wind above the disk that is ionized by energetic
photons from the central regions of the system.  If this is the case,
the constraints shown in Figure~\ref{recomb10000} apply to the ionized
atmosphere or wind, not the neutral core of the disk.  If a wind is
involved \citep[see, for example][]{K94}, it is likely that the the
velocity would be outside the observed window, providing a natural
explanation for the lack of an observed recombination line even if it
weren't for the effects of the external radiation.

\section{Summary}

A search has been made for the H65$\alpha$ recombination line in the
radio source 3C~84 in the center of NGC1275.  Observations of
free-free absorption against the northern feature at about 2.5 pc
projected distance from the core of this source suggested that, for a
wide range of possible densities, such recombination lines could be
observed.  The recombination lines were only expected to be observed
against the weak northern feature which is probably the receeding jet
on the far side of the source.  The ability of the VLBA to separate
this feature from the rest of the source spatially was used to relax
the bandpass calibration requirements and to allow the use of a novel
scheme to to do an accurate bandpass calibration based on the source
itself.  These advantages allowed a much deeper search for the
recombination line to be made than would have been possible with lower
spatial resolution.

No H65$\alpha$ line was found.  The limit on the optical depth is
approximatly 0.15, depending on exactly how the measurements are made
and what region is measured.

\begin{figure} 
\epsscale{0.70} 
\plotone{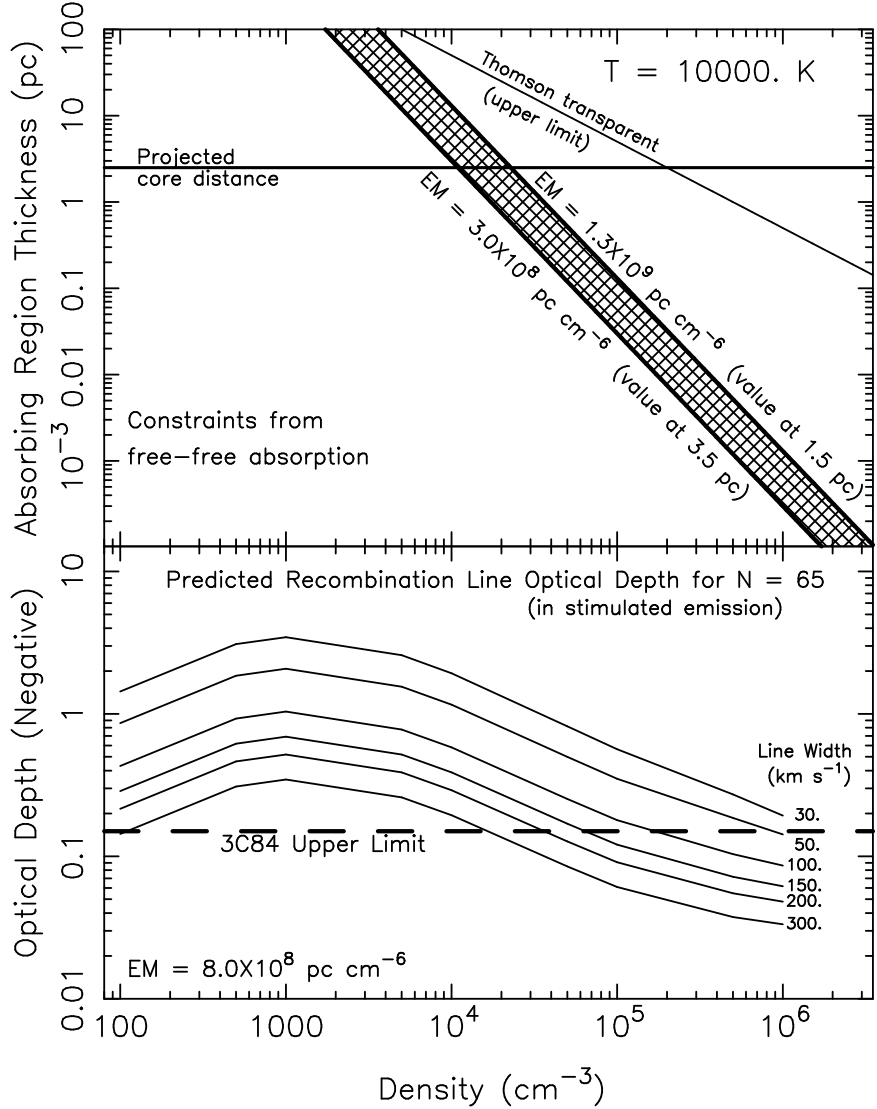} 
\caption{Constraints placed by the observations on the physical
parameters of the free-free absorbing material, assuming an electron
temperature of 10$^4$ K.  The top panel shows the thickness of the
absorbing material as a function of density based on the measured
emission measure.  A band of emission measures is shown representing
the range observed over the face of the counterjet by \citet{W00}.
Size limits due to Thomson scattering and to the overall scale of the
region are also indicated.  The lower panel shows the predicted
H65$\alpha$ recombination line optical depth in a benign radiation
environment as a function of density and line width, based on an
emission measure in the middle of the observed range.  The lack of a
detection would constrain the absorbing material to have conditions
that fall in the region below the indicated line or to have a velocity
outside the observing window.  However the radiation environment is
not benign.  Saturation and radiation damping caused by the jet radiation
are expected to make the the line unobservable.
\label{recomb10000}}
\end{figure}

The non-detection is expected because of saturation and radiation
damping caused by the intense radio radiation from the jets.
Therefore conclusions cannot be made about the physical conditions in
the the free-free absorbing region unless that region is actually much
further from the jets than the geometry would suggest --- a bare
minimum of twice as far as expected and probably much further.  If
saturation and radiation damping were not an issue, the recombination
line optical depth limit would constrain the free-free absorbing
material to have a combination of high line width and high density, as
shown in Figure~\ref{recomb10000}.  Alternatively, the absorbing
material could have a velocity significantly offset from the systemic
value.  High line widths and offset velocities are often seen in AGN,
including NGC~1275.  The most extreme velocities and line widths are
in the BLR, which is smaller than the region of the absorption, but
velocities adequate to hide the recombination line could occur on the
appropriate scales.

\acknowledgements 

Sadly, Anantharamaiah succumbed to cancer in October 2001 before this
paper was finished.  He was well loved by all who knew him and he is
sorely missed.  We are grateful to Miller Goss for a critical and
helpful reading of the manuscript and to Joan Wrobel for both reading
the manuscript and for provoking us to seek out a discussion of the
effects of the radiation field on the maser emission.

\clearpage


\end{document}